\documentclass[twocolumn, aps]{revtex4}
\usepackage{natbib}
\usepackage{amsmath}

\usepackage{graphicx}

\begin{document}

\title{Protein Structure Prediction: The Next Generation}
\author{Michael C. Prentiss}

\affiliation{Center for Theoretical Biological Physics, La Jolla, CA 92093, Department of Chemistry and Biochemistry,  University of California at San Diego, La Jolla, CA 92093
}

\author{Corey Hardin}
\affiliation{Department of Chemistry, University of Illinois at Urbana-Champaign, 600 South Mathews Ave, Urbana, IL 61801
Urbana, IL 61801}

\author{Michael P. Eastwood}
\affiliation{Department of Chemistry and Biochemistry,  University of California at San Diego, La Jolla, CA 92093
}

\author{Chenghong Zong}
\affiliation{Center for Theoretical Biological Physics, La Jolla, CA 92093, Department of Chemistry and Biochemistry,  University of California at San Diego, La Jolla, CA 92093
}
\author{Peter G. Wolynes}
\affiliation{Center for Theoretical Biological Physics, La Jolla, CA 92093, Department of Chemistry and Biochemistry,  Department of Physics University of
    California at San Diego, La Jolla, CA 92093}

\date{\today}

\begin{abstract}
Over the last 10-15 years a general understanding of the chemical
reaction of protein folding has emerged from statistical mechanics.
The lessons learned from protein folding kinetics based on energy
landscape ideas have benefited protein structure prediction, in
particular the development of coarse grained models. We survey results
from blind structure prediction.  We explore how second generation
prediction energy functions can be developed by introducing
information from an ensemble of previously simulated structures.  This
procedure relies on the assumption of a funnelled energy landscape
keeping with the principle of minimal frustration.  First generation
simulated structures provide an improved input for associative memory
energy functions in comparison to the experimental protein structures
chosen on the basis of sequence alignment.
\end{abstract}

\maketitle
Every other summer, research groups compare their different protein
structure prediction methods via the Critical Assessment of Techniques
for Protein Structure Prediction (CASP) experiment.  During the CASP
experiment, sequences of experimentally determined protein structures
that are not public available are placed on the web. This exercise is
double blind where neither the organisers nor the participants know
the experimentally determined structure.  Groups respond with up to 5
ranked predictions, before a predetermined date, such as the 
publication of the structures.  Since the inception of CASP, three
dimensional structure prediction category has expanded to address
related prediction questions such as sequence to structure alignment
quality, amino acid sidechain placement, multi-domain domain
boundaries, and the ordered or disordered nature of a protein sequence
\cite{Moult}.

These different prediction questions can be examined from a common
framework: the principle of minimal frustration.  The principle of
minimal frustration states that native contacts must be more
favourable, in a strict statistical sense \cite{GoldsteinRA-AMH-92},
than non-native contacts in order for proteins to fold on physiologic
time scales \cite{BryngelsonJD87}.  Without a sufficient energetic
bias towards the native state, the multi-dimensional energy surface as
a function of native structure possesses too many minima for an
efficient stochastic search.  Such an energy surface would lead to
slow folding kinetics, even if the proteins ever found a sufficiently
stable native state.  This is not true since we know most proteins
fold without assistance \cite{AnfinsenCB73}.  The opposite of a
rough energy surface is biased toward the native basin without any
local minima is an absolute manifestation of the principle of minimal
frustration. Funnelled energy surfaces have no unfavourable
energetic traps (\emph{i.e.} G\=o Models) have been shown to reproduce
most features of experimental folding kinetics \cite{GoN83,
Koga_Takada01, portman98}. These energy landscape concepts can richly
be applied in several areas of chemistry and physics\cite{Wales}.
Apparently, evolution's energy function is minimally frustrated.

\begin{figure}{\par\centering
 \resizebox*{3.0in}{2.0in}{\includegraphics{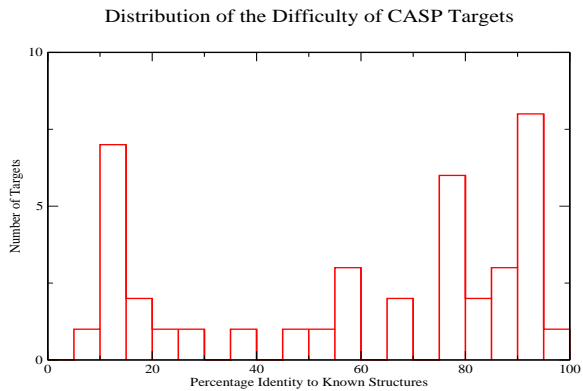}}
 \par}
\caption{\label{casp6_difficulty}The difficultly of the prediction
 targets as defined by percent identity. Proteins below 25\%
 sequence identity are usually considered \textit{ab initio} or
 fold recognition targets.}
\end{figure}

The correlation between a protein sequence and its three dimensional
structure can be described using similar landscape language.  As a
protein sequence diverges away from a consensus wild type sequence,
the potential for energetically unfavourable interactions increases.
The wild type sequence and its homologues will fold toward the same
native basin.  Only once enough frustrating contacts are added to the
wild type sequence will the sequences no longer correspond to the native
state ensemble.  Sequences with over 25\% sequence
identity to previously determined protein structures are called
comparative modelling targets.  The energy landscape underlying such a
prediction is a G\=o model based on the structure of the known
homologue. This heavily funnelled energy surface yields
high resolution structures, with the discrepancies the turns and
residues which have poor sequence to structure alignments.
Fig.~\ref{casp6_difficulty} demonstrates the distribution of homology
of proteins sequence to known structures included in CASP6.  Since
proteins below 25\% sequence identity are considered new fold
recognition targets, 70\% of the structures were comparative modelling
targets.  Recently sequenced genomes such as \textit{E. coli} have the
same ratio of \textit{ab initio} to comparative modelling targets,
which suggests the analysis of this ratio over time could be a useful
measure of the progress of efforts to experimentally find examples of
all of Nature's protein structures. 
 
\begin{figure}{\par\centering
 \resizebox*{3.0in}{2.0in}{\includegraphics{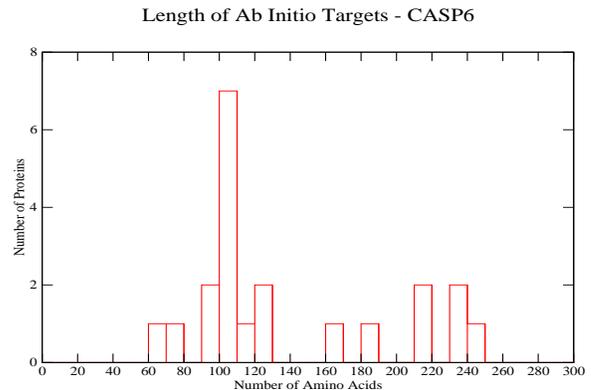}}
 \par}
 \caption{\label{casp6_ab_length}The \textit{ab initio} prediction
     targets amino acid lengths for CASP6.}
\end{figure}

 In contrast to comparative modelling, \textit{ab initio} structure
 predictions do not have the advantage of creating G\=o like energy
 surfaces.  While many \textit{ab initio} targets contain less than 150
 residues, and thus are candidates for standard techniques, there are
 several that are longer as shown in Fig.~\ref{casp6_ab_length}.  Most
 longer sequences will be multi-domain proteins. This causes new
 problems.  Folding a protein with two hydrophobic cores allows
 for new sources of frustration, beyond those present in single domain
 proteins.  To obtain predictions for such problematic sequences, they
 usually must be divided into their constituent domains.  Current
 methods for dividing the sequence into domains range from purely
 sequence based algorithms, which look for sequence patterns in
 multiple sequence alignments, to simulation techniques that look for
 hydrophobic core formation amongst multiple independent simulations
 \cite{Wheelan_etal00,Heringa02,Rigden02}.

\begin{figure}
     \centering \includegraphics[width=\linewidth]{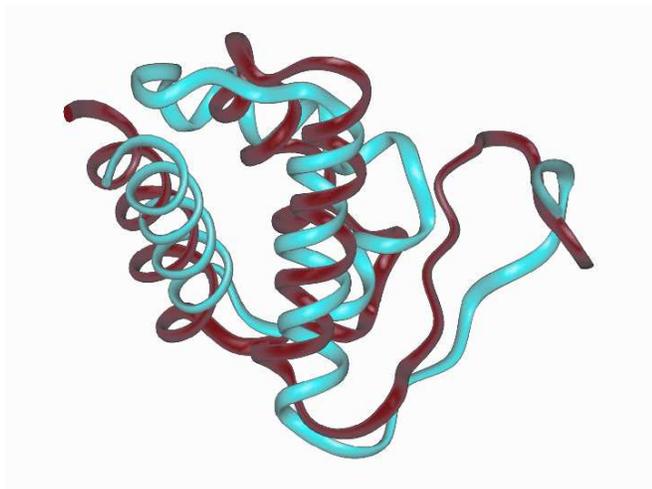}
     \caption{\label{casp5t0170overlay}Sequence dependent
       superpositions of Model 1 structure against the native state for
       CASP5 target T0170 (PDB ID 1UZC).  Blue represents the
       prediction and the native state is represented with red.}
 \end{figure}

\begin{figure}{\par\centering
 \resizebox*{2.0in}{2.0in}{\rotatebox{90}{\includegraphics{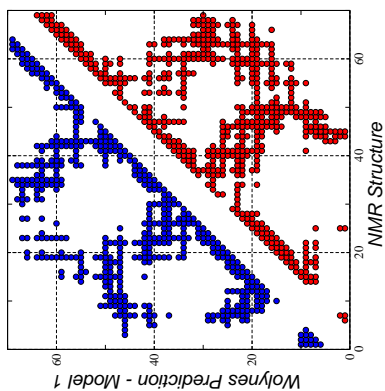}}}
 \par}
 \caption{\label{casp5t0170contactmap}Contact map of target T0170 (PDB
     ID 1UZC) model
   1 structure against the NMR structure.}
 \end{figure}

 The case studies we highlight of difficult structure predictions were
  chosen from our participation in the CASP5 and CASP6 experiments.  In
  CASP5, we utilised several improved techniques, such as a backbone hydrogen
  bond term for the proper formation of beta sheets, and a liquid
  crystal like term to ensure parallel or anti-parallel sheet formation
  \cite{Hardin2002ab}.  We also performed target sequence averaging
  which enhances the funnelling of the prediction landscape
  \cite{Eastwood00}, and assessed our ensemble of sampled structures
  with a twenty letter contact for submission \cite{KoretkeKK96}.  Our
  most striking result from this round of blind prediction was a
  prediction for target T0170 protein databank \cite{Berman} code (PDB
  ID IUZC).  Fig.~\ref{casp5t0170overlay} presents the sequence dependent
  overlay of our Model 1 structure with the experimentally determined
  structure.  The sequence dependent alignment quality of this structure
  is high as measured by a Q score of 0.38. Q is an order parameter
  defined in Eq.~\ref{q} that measures the sequence dependent structural
  complementarity of two structures, where Q is defined as a normalised
  summation of C-alpha pairwise contact differences.
   \begin{equation}
   \label{q}
   Q=\frac{2}{(N-1)(N-2)}\sum_{i<j-1}
       \exp\left[-\frac{(r_{ij}-r_{ij}^{\rm N})^2}{\sigma_{ij}^2}\right]
  \end{equation}      
  The resulting order parameter, Q, ranges from 0, when there is no
  similarity between structures at a pair level, to 1 which is an exact
  match.  Q has been shown to be more sensitive in determining the
  quality of intermediate quality protein structure predictions
  \cite{Eastwood00}. Q scores of 0.4 for single domain proteins equals an
  RMSD of 5$\text{\AA}$.  In most cases the reference state for the Q
  score is the native state, but often one wants to compare structural
  similarity between structures in a simulation.  A sequence independent
  measure CE\cite{Bourne98}, also scores well(CE Z-score = 4.1).  The CE
  Z-score measures structural complementarity without regard to sequence
  information, and is parameterised such that structures between with a
  Z-Score greater than 4 belong to the same protein
  structure family.  The contact map of the prediction,
  Fig.~\ref{casp5t0170contactmap} which identifies all of the C-alpha
  intermolecular interactions within 9$\text{\AA}$ where the axes 
  are the index of the protein, shows the correct packing of the helices.

\begin{figure}
      \centering
      \includegraphics[width=\linewidth]{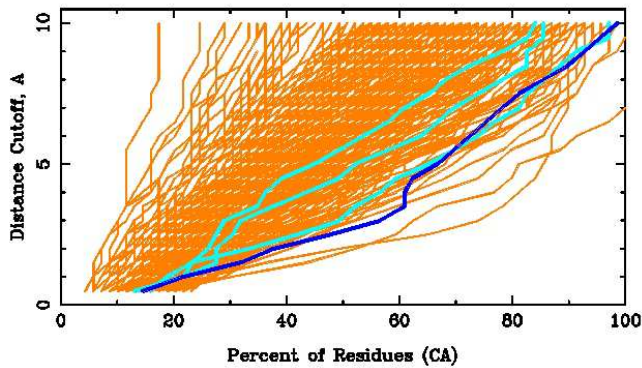}
      \caption{\label{casp6_hubbard}Percentage of residues under a RMSD
        limit. (Dark Blue - Model 1, Light Blue - Model 2-5, Orange - Other
        Groups Prediction)}
  \end{figure}

  Fig.~\ref{casp6_hubbard} shows the size of partially correct
  continuous in sequence segments under an RMSD cutoff.  When compared
  against the other predictions, our Model 1 prediction (Dark Blue) was
  amongst the best of all submitted structures.  Also the relative
  success of the prediction, classifies this target as being of
  moderate difficulty.  In this example CASP demonstrates small (70
  residues) all alpha proteins are beginning to be successfully
  predicted by a variety of \textit{ab initio} techniques.

 \section*{Methods}

 \subsection*{Energy Functions and Sampling}

 We used an Associative Memory Hamiltonian (AMH), with optimised
 parameters to sample and predict structures
 \cite{FriedrichsMS89,FriedrichsMS90,FriedrichsMS91}. The AMH uses a
 reduced description of the amino acid chain in order to gain the
 orders of magnitude computational acceleration over all atom models
 needed to fold moderate length proteins with ordinary
 computational resources, and has been described in great detail before
 \cite{Eastwood00}.  This is possible due to reducing the number of
 atoms per residue from over 10 to only three backbone atoms: the
 \(C_{\alpha},C_{\beta}\), and \(O\). The remaining backbone heavy
 atoms (\(N,C'\)) can be reconstituted using the ideal geometry of the
 peptide bond as a template. Also we reduced the complexity of the
 amino acid code from twenty letters, to four.  We chose the four
 letter code, which has the advantage of preserving a diversity of
 contacts, because it is still simple enough that the number of
 coefficients that need to be optimised does not create problems of
 inaccurate statistics due to limits of interactions encountered in the
 molten globule state.  Specifically four amino acid classes are
 defined: hydrophilic (A, G, P, S, T), hydrophobic (C, I, L, M, F, W,
 Y, V), acidic (N,D,Q,E), and basic (R,H,K) \cite{Hardin00}. The
 optimisation procedure produces an energy landscape that discriminates
 the native state from misfolded states, while avoiding kinetic traps
 reasonably well \cite{GoldsteinRA92,GoldsteinRA-AMH-92}.  The AMH is
 an analogue to the neural networks designed by Hopfield to synthesise
 information from multiple previous experiences \cite{Hopfield_1982}.
 This energy function recalls structural patterns in a set of known
 protein structures.  The Hamiltonian produces an energetically
 favourable minimum when there is sufficient coherence between a set of
 three dimensional protein structures.

 The AMH energy function, in its most general sense, consists of a
 backbone term, $E_{\rm back}$ and interaction term, $E_{\rm int}$
 defined by,
 \begin{equation}
   \label{amh_function}
   E_{\rm total}=E_{\rm back}+E_{\rm int}.
 \end{equation} 
 The backbone energy term consists of several terms that reproduce the
 self-avoiding behaviour of the polypeptide chain give by,
 \begin{equation}
   \label{amh_back}
   E_{\rm back}=-(E_{\rm SHAKE}+ E_{\rm rama} + E_{\rm ev} + E_{\rm chain} + E_{\rm chi}).
 \end{equation}
 As in many molecular mechanics energy functions, covalent bonds are
 preserved by using the SHAKE algorithm \cite{Ryckaert77} \(E_{\rm
   SHAKE}\), which enables an increase of the time step size, and
 eliminates the need for a traditional harmonic calculation.  The SHAKE
 algorithm preserves the distances between neighbouring
 \(C_{\alpha}\)-\(C_{\beta}\), and \(C_{\alpha}\)-\(O\) atoms.  The
 neighbouring residues limit the variety of angles the backbone atoms
 can occupy, producing a Ramachadran plot \cite{Rama}.  This
 distribution of angles is reinforced by a potential, \(E_{\rm rama}\)
 with low barriers to encourage rapid local backbone movements.
 Another term, \(E_{\rm ev}\) maintains a sequence specific excluded
 volume constraint between \(C_{\alpha}\)-\(C_{\alpha}\),
 \(C_{\beta}\)-\(C_{\beta}\), \(O\)-\(O\), \(C_{\alpha}\)-\(C_{\beta}\)
 atoms.  The chain connectivity, and planarity of the peptide bond due
 to resonance is ensured by means of a harmonic potential, \(E_{\rm
   chain}\).  Also the chirality of the \(C_{\alpha}\), due to its four
 different bonding partners is maintained using scalar product of
 neighbouring unit vectors of carbon and nitrogen bonds, \(E_{\rm
   chi}\).

 While \(E_{\rm back}\) creates peptide like stereo-chemistry, it 
 does not introduce the majority of the attractive
 interactions that result in folding. Such interaction are supplied by
 the rest of the potential \(E_{\rm int}\). The interactions described
 by $E_{\rm int}$ depends on the sequence separation $\left\vert i-j
 \right\vert$. Specifically,
 they are divided into three proximity classes $x(\left\vert
   i-j\right\vert)$: $x={\rm short}$ ($\left\vert i-j\right\vert<5$),
 $x={\rm medium}$ ($5\le\left\vert i-j\right\vert\le12$) and $x={\rm
   long}$ ($\left\vert i-j\right\vert>12$) as defined by Eq.~\ref{eint}. 
 \begin{equation}
 \label{eint}
 E_{\rm int}=E_{\rm short}+E_{\rm med}+E_{\rm long}.
 \end{equation}
 Also these distance classes are also referred to as local, 
 super-secondary, and tertiary, respectively.

 The AMH interaction potential \(E_{\rm int}\) is based on correlations
 between a target's sequence signified by \(i,j\), and the
 sequence-structure patterns in a set of memory proteins \(\mu\)
 represented as \(i',j'\), and a pairwise contact potential.  The pairs
 in the target and in the memory are first associated using a
 sequence-structure threading algorithm \cite{KoretkeKK96}. The
 database is assumed to contain a subset of pair distances, which may
 match the associated pair distances in the target structure.  The
 general form of the associative memory interaction is:
  \begin{equation}
  \begin{split} 
   \label{amh_general}
    E_{\rm int}&= -\frac{\epsilon}{a}\sum ^{N_{mem}}_{\mu}\sum_{j-12 \leq i \leq j-3} \\             
     & \gamma(P_{i},P_{j},P^{\mu}_{i'},P^{\mu }_{j'})
   \exp\left[-\frac{(r_{ij}-r_{i'j'}^{\rm
      \mu})^2}{2\sigma^2_{ij}}\right] \\
           &+ -\frac{\epsilon}{a}\sum_{k=1}^{3}C_{k}(N)\gamma (P_{i},P_{j},k)U_{k}(r_{ij})
  \end{split}
  \end{equation}
 where the similarity between target pair distances \(r_{ij}\), with
 aligned memory pair distances \(r^{\mu}_{i'j'}\) is measured by Gaussian
 functions whose widths are given by \(\sigma_{ij}=\left\vert
   i-j\right\vert^{0.15}\text{\AA}\). The set of parameters,
 \(\gamma\), encode the similarity between residues i and j, and the
 memories residues i' and j'.  Favourable interactions occur during
 coherence in the distances achieved in the sequence to structure
 alignments. The encoding of the alignment information in
 Eq.~\ref{amh_general} is only an example of what is used for the
 all-alpha energy functions.  Other encodings have been used in the
 alpha-beta energy function \cite{Hardin2002ab} to improve the
 discrimination between helices and strands. While the first term in
 Eq.~\ref{amh_general} is the superposition of interactions over a set
 of experimentally determined structures, it also shares a dependence
 on the sequence separation between the interacting residues.  For
 residues separated by greater than 12 residues, a contact potential
 \(E_{\rm long}\), as described by the second term in
 Eq.~\ref{amh_general}, which does not depend on interaction
 information from the structures used to define local in sequence
 interactions.  In this term $C_{k}(N)$ represents a sequence length
 dependence scaling to account for the variation in probability
 distributions based on sequence length. Five wells instead of the
 three defined here by $U_{k}(r_{ij})$ determine interactions in the
 alpha-beta energy function \cite{Hardin2002ab}.  Energy units
 $\epsilon$ are defined excluding backbone contributions in terms of a
 native state energy in Eq.~\ref{units},
 \begin{equation}
   \label{units}
   \epsilon=\frac{\left\vert E^{\rm N}_{\rm amh}\right\vert}{4N},
 \end{equation}
 where $N$ is the number of residues. A distance class scaling $a$, is
 constant in each of the energy classes because they are designed to
 be equal during the optimisation 

 The solvent in these energy functions is treated in a mean field
 manner, where the implicitly solvated native states of the proteins define the
 energy gap to the molten globule state.  Solvent effects are also
 present in the sequence to structure alignment energy functions, but
 they are not explicitly represented in the molecular dynamics energy
 function. Water mediated contacts with an expanded 20 letter code in
 the contact potential were introduced \cite{Papoian04pnas}, based upon
 previous work which examined protein recognition
 \cite{Papoian03biopoly, Papoian03jacs}.  The water mediated contacts
 along with a new one dimensional burial term has shown promising
 results especially for long proteins.

 Once the energy function is optimised, the minima of the energy
 function are probed via simulated annealing with molecular dynamics
 simulations.  This minimisation technique integrates Newton's
 equations of motions to determine the energy of the next time step.
 Simulated annealing slowly reduces the temperature from a high value
 as in the tempering of steel in metallurgy.  This minimisation algorithm 
 allows for local searches, while allowing modest energy barriers 
 to be overcome.

 Energy landscape ideas have generated an optimisation scheme for
 creating funnelled energy surfaces.  While funnelled, the
 parameterisation does not eliminate all non-native minima.  The
 superposition of several energy surfaces reduces the likelihood of
 such trapping in local minima \cite{maxfield79,finkelstein98}.  The
 flexibility of the AMH framework provides several ways of
 incorporating multiple sequence alignment information.  Some of the
 options include creating a consensus sequence \cite{Eastwood00},
 simulating different homologue sequences concurrently, and averaging
 the resulting forces and energies \cite{Hardin2002ab}.  The averaged
 AMH energy function we used average the forces and the energies of
 these simulation over a set of sequences, because it allows for more
 generalisable results than may occur with other techniques, and is
 described as in Eqs.~\ref{msa1},~\ref{msa2} ,

  \begin{equation}
   \begin{split}
   \label{msa1}
    E_{\rm short + medium} &= -\frac{1}{N_{seq}}\frac{\epsilon}{a} \sum ^{seq}_{1
      }\sum ^{N_{mem}}_{\mu}\sum_{j-12 \leq i \leq j-3} \\ 
       & \gamma (P_{i},P_{j},P^{\mu}_{i'},P^{\mu }_{j'} \exp\left[-\frac{(r_{ij}-r_{i'j'}^{\rm \mu})^2}{2\sigma^2_{ij}}\right] 
  \end{split}
  \end{equation}

 \begin{equation}
  \label{msa2}
    E_{\rm long} = -1/N_{seq}\frac{\epsilon}{a} \sum ^{seq}_{1 }\sum_{k=1}^{3}C_{k}(N)\gamma (P_{i},P_{j},k)U_{k}(r_{ij})
 \end{equation}

\begin{figure}{\par\centering
 \includegraphics[width=\linewidth]{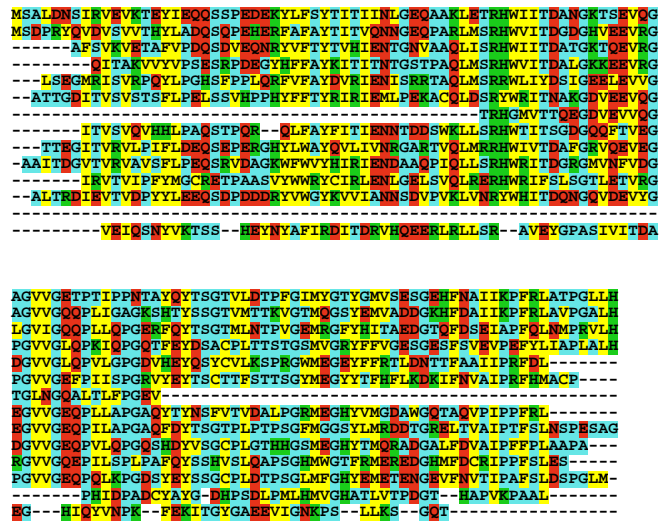}
 \par}
 \caption{\label{casp6t0212_msa} Multiple sequence alignment for target
   T0212 (PDB 1TZA) coloured with respect to a four letter code, where red
   represents acidic residues, blue represents polar residues, yellow
   represents nonpolar residues, and green represents basic residues.}
 \end{figure}

 To superimpose multiple energy landscapes, we need a multiple sequence
 alignment to a set of sequence homologue.  Sequences homologous to
 the target sequence are first identified by using PSI-Blast with
 default parameters \cite{Altschul1997}.  Each sequence above and below
 a certain sequence identity thresholds (70\% 30\% in this work) is
 then aligned against each other, and proteins that have greater than
 90\% sequence identity to other identified sequence homologues are
 removed.  The culling of the sequence homologues via open source
 bioinformatic libraries is necessary for two reasons \cite{bioperl}.
 Some classes of proteins have a large number of sequence homologues,
 and performing a multiple sequence alignment can be impractical.  Also
 removing sequence homologues attempts to remove biases introduced when
 there are few homologues.  The remaining sequences were aligned using
 a multiple sequence alignment algorithm\cite{CLUSTAL}.  Within the AMH
 energy function, gaps occurring in a sequence alignment could be
 addressed in a variety of ways, in this work gaps in the target
 sequence are ignored, while gaps within homologues are completed with
 residues from the target protein.  This strategy may introduce biases
 toward the target sequence, but this approach is preferred to perhaps
 ignoring interactions.  Fig.~\ref{casp6t0212_msa} shows a
 representative multiple sequence alignment for a target, coloured with
 respect to the four letter code of the AMH.  If one focuses on the
 hydrophobic yellow residues, the alternating hydrophobic hydrophilic
 patterns for beta strands formation are apparent.

 Another way of introducing the characteristics of multiple funnelled
 energy landscapes is using information derived from neural networks
 trained on multiple sequence alignments.  Even with different
 architectures, neural networks typically achieve 75\% accuracy when
 predicting secondary structure.  Recently it has been shown artful
 combinations of two different predictions can slightly improve the
 results \cite{Zhang_etal03}.  This secondary structure information was
 added by a biasing energy function to either a helix or a strand via,
 $E_{Q_{ss}} = 10^5 \epsilon(Q-Q_{ss})^4$ \cite{Eastwood00}, where
 ${Q_{ss}}$ is defined by Eq.~\ref{qmf_ss},
 \begin{equation}
   \label{qmf_ss}
   Q_{\rm ss}= \sum_{k}^{n}\frac{2}{(N_{k}-1)(N_{k}-2)}\sum_{i<j-1}\exp\left[-
     \frac{(r_{ij}-r_{ij}^{\rm ss})^2}{\sigma_{ij}^2}\right].
 \end{equation} 
 ${Q_{\rm ss}}$ is takes the same form of the $Q$ define before in
 Eq.~\ref{q} except that potential acts over $n$ independent secondary
 structures units derived from secondary structure prediction. The
 distances that define energy minimum, $r_{ij}^{\rm ss}$ are determined
 from experimentally determined Cartesian distances.  Previously in an
 effort to incorporate this secondary structure information, the
 Ramachandran potential has been altered to bias the backbone
 \cite{Hardin02a}.  The local in sequence potential $E_{\rm Q_{\rm
     ss}}$ is preferred to the Ramachandran potential biasing because
 it avoids SHAKE violations when the strength of the bias is increased.

 For most selected CASP6 targets, we followed the same protocol.  We
 averaged the AMH potential over multiple sequence homologues when they
 were available.  In most cases, information from secondary structure
 prediction was used to bias secondary structure units to their
 predicted structures.  Molecular dynamics with simulated annealing 
 sampled low energy structures.  Also constant temperature
 slightly above the predicted glass temperature were used to generate
 candidate structures. We collected structures above \(T_{K}\), which
 usually gives the fastest folding thereby compromising between the
 funnelled and glassy behaviour of the energy function.  Once the
 kinetics of the structure slows, the diversity of structures
 encountered disappears.  The slow kinetics regime typically
 predominates around a temperature of 0.9.  While using a linear
 annealing schedule up to \(T_{K}\), about 25 different collapsed
 structures were collected during each simulation.  The amount of
 sampling performed for each structure varied from about 500 to 20,000
 different structures.  While this was roughly 50 times more sampling
 than we had previously performed in the CASP setting, it is dwarfed by
 the efforts of others who can sample in the millions of structures by
 using more powerful computational resources \cite{BONN2001b}.
 Subsequently, a smaller subset of structures was selected for
 submission by evaluating the size of the hydrophobic core and the
 hydrophilic surface area.  Further selection criteria included visual
 inspection, agreement with the preliminary secondary structure
 prediction, and low energies predicted from a second optimised contact
 energy function.

 \subsection*{Selection of Structures}

 To select candidate structures from independent simulated annealing or
 constant temperature trajectories, we calculated both the buried
 hydrophobic surface area and the exposed hydrophilic surface area
 along the trajectory.  In an effort to calculate the buried or exposed
 surface area, we assigned residues which have greater than the mean
 total surface area as solvent exposed, and the converse as solvent
 buried.  We scaled each surface area by a weight to represent the
 likelihood of amino acid burial. It was modelled to the free energy
 cost of transferring each amino acid from octanol to water
 \cite{zhou:zhou04} in an effort to introduce a sequence specificity
 as shown in Eq.~\ref{sa},
      \begin{equation} 
       \label{sa}
        E_{\rm Burial}=\sum_{i}^{N}
          \begin{cases}
           \gamma_{i}*SA_{i},& \text{if $ SA_{i} > $ total surface} \\ 
                          0,&  \text{if $ SA_{i} > $ total surface}
          \end{cases}
      \end{equation}

This normalisation is desirable because the surface accessibility is
  calculated from our minimal \(C_{\alpha}, C_{\beta }\), and \(O\)
  atoms, which produces amino acids of the same volume.  Such an energy 
 term would be more valuable if non-additive interactions, and a larger 
 number of hydration layers were added.  The
  unavoidable inaccuracies in atomistic force fields, and the slow
  glassy kinetics of sidechain rearrangements prevented any completion
  of the backbone and sidechains with all-atoms or minimisation of
  putative structures \cite{kussell:168101}.

  \begin{table}
    \caption{\label{table1} Linear Regression of Hydrophobic Burial Energy }
    {\centering \begin{tabular}{|c|c|c|} \hline Proteins & fold class &
        Correlation  Coefficient \\ \hline \hline
        1R69&\(\alpha\)&.22 \\
        1BG8&\(\alpha\)&.33 \\
        1UTG&\(\alpha\)&.63 \\
        1MBA&\(\alpha\)&.40 \\
        2MHR&\(\alpha\)&.46 \\
        1IGD&\(\alpha/\beta\)&-.70 \\ 
        3IL8&\(\alpha/\beta\)&-.06 \\
        1TIG&\(\alpha/\beta\)&.02    \\ 
        1BFG&\(\beta\)&.16 \\
        1CKA&\(\beta\)&-.14 \\
        1JV5&\(\beta\)&.11 \\
        1K0S&\(\beta\)&.27  \\
        \hline
      \end{tabular}\par}
  \end{table}


  Another parameter we used after sampling to select and examine
  structures was based on sequence specific backbone probabilities.  The
  specificity of local interactions have been fruitful for improving
  collapsed proteins structure predictions \cite{Baker97}.  In a similar
  spirit sequence specific nearest neighbour probabilities were also used
  \cite{Betancourt:2004}.  Local signals have also been theoretically
  shown to contribute roughly a third of the total folding gap for
  $\alpha$ helical proteins \cite{SavenJG96}.  Similarly we started
  looking at such probabilities to further improve the backbone
  potential of the AMH, but without needing secondary structure
  prediction.
  \begin{equation} 
   \label{mscore}
    E_{\rm trimer}=\sum_{i=2}^{N-1}Log P(i-1,i,i+1,\phi,\psi)
  \end{equation}
  Somewhat surprisingly, the summation of the resulting $\log$
  probabilities from 4,012 highly resolved protein structures could be
  used as an additional measure as part of a strategy for the selection
  of structures out of an ensemble.  Table~\ref{table1} shows the linear
  correlation coefficients between structures of varying Q-scores,
  sampled above \(T_{\rm K}\) which is where the best predictions
  usually occur before glassy dynamics dominates the kinetics.  For both
  proteins with all \(\alpha\), and \(\alpha/\beta\) compositions, the
  summed log probabilities provide discrimination, but not within the
  all \(\beta\) folds.  These results shown in Table~\ref{table_marcio}
  echo the previous findings in terms of the \(\phi\), \(\psi\)
  probability maps and also that all beta structures are less well
  predicted when a dihedral angle energy function is minimised.  The
  weakness of nearest neighbour excluded volume effects to determine
  local structure is also demonstrated in the consistent weakness of
  secondary structure prediction with respect to beta strands. Alpha
  helices are correctly predicted to roughly 80\% accuracy while beta
  strands average 60\% accuracy by such pure sequence based algorithms.
  The difficulty of predicting some circular dichroism spectroscopy
  results for beta to coil transitions can also be attributed to the
  weakness of the local backbone excluded volume interactions.

  \begin{table}
    \caption{\label{table_marcio} Linear Regression of Mscore }
    {\centering \begin{tabular}{|c|c|c|} \hline Proteins & fold class &
        Correlation  Coefficient \\ \hline \hline
        1R69&\(\alpha\)&.29 \\
        1BG8&\(\alpha\)&.04 \\
        1UTG&\(\alpha\)&.26 \\
        1MBA&\(\alpha\)&.26 \\
        2MHR&\(\alpha\)&.10 \\
        1IGD&\(\alpha/\beta\)&.37 \\
        3IL8&\(\alpha/\beta\)&.13 \\
        1TIG&\(\alpha/\beta\)&.19 \\
        1BFG&\(\beta\)&.08 \\
        1CKA&\(\beta\)&.03 \\
        1JV5&\(\beta\)&-.07 \\
        1K0S&\(\beta\)&-.10 \\
        \hline
      \end{tabular}\par}
  \end{table}


  \section*{Results}
  \subsection*{Blind Simulations}

  For \textit{ab initio} blind predictions in CASP6, we selected
  sequences if there were no experimentally determined homologous
  structures found by automated comparative modelling servers.  The
  overall results for the \textit{ab initio} structure prediction
  simulation are summarised in Table~\ref{casp6_table1}, where the
  abbreviations are length = the number of amino acids, temp =
  temperature where best structure was encountered, sub Q or samp Q =
  the best sampled and submitted structures respectively as a judged 
  by a function of Q, and traj = number of independent trajectories 
  simulated.  The
  CASP6 targets are classified under the following categories (NF=new
  fold, FR/A=fold recognition analog, FR/H=fold recognition homologue,
  CM/H=comparative modelling hard).  Targets T0207, and T0270 where
  removed from the experiment so their CASP class are undefined.  Structures for T0207 and
      T0272-b were not submitted.  There
  are a few main points from this data.  Using a Q of 0.4 as a measure
  successful prediction, we were able to encounter high quality
  structures for 4 targets and nearly so for 4 others.  The temperature
  at which the best structures were sampled was between the 1.2 and 0.8,
  which is the annealing regime we investigated most throughly.  This
  suggests our annealing schedules were close to the behaviour we sought
  \textit{a priori}. The longer the length of the target sequence
  clearly reduced the quality of our predictions.  Also the proteins
  where we had a greater number of trajectories naturally showed better
  structures. A final observation identifies the difference between the
  best submitted structure and the best sampled structure as
  disappointingly large for some of the targets.  This can be attributed
  our strategy of maximising the number of simulations performed rather
  than more carefully studying our trajectories.  This difference would
  be smaller if greater care was taken in the selection of the
  structures, but the number of high quality structures would have been
  less.
  \begin{table}
    \caption{\label{casp6_table1}CASP6 Results: Best Submitted and Sampled Structures  }
    {\centering \begin{tabular}{|c|c|c|c|c|c|c|c|} \hline 
        target & length & fold &sub Q&samp Q& temp & traj & CASP \\ \hline \hline
        T0281   & 70  & \(\alpha/\beta\) &.34 &.48 & 0.85 & 986 & NF     \\
        T0201   & 94  & \(\alpha/\beta\) &.36 &.44 & 1.39 & 199 & NF     \\ 
        T0212   & 123 & \(\beta\)        &.26 &.42 & 1.30 &  97 & FR/A   \\ 
        T0230   & 102 & \(\alpha/\beta\) &.31 &.42 & 1.05 & 395 & FR/A   \\
        \hline   \hline
        T0207   & 76  & \(\alpha/\beta\) & -- &.39 & 0.98 & 297 & --     \\
        T0224   & 87  & \(\alpha/\beta\) &.30 &.38 & 1.20 & 501 & FR/H   \\
        T0263   & 97  & \(\alpha/\beta\) &.34 &.38 & 0.94 & 404 & FR/H   \\ 
        T0272-a & 85  & \(\alpha/\beta\) &.30 &.37 & 0.94 &  30 & FR/A   \\ 
        T0265   & 102 & \(\alpha/\beta\) &.29 &.34 & 0.83 & 374 & CM/H   \\
        T0213   & 103 & \(\alpha/\beta\) &.26 &.32 & 0.98 & 448 & FR/H   \\
        T0243   & 88  & \(\alpha/\beta\) &.31 &.32 & 0.95 & 418 & FR/H   \\
        T0239   & 98  & \(\alpha/\beta\) &.25 &.32 & 0.99 & 424 & FR/A   \\ 
        T0214   & 110 & \(\alpha/\beta\) &.24 &.30 & 0.41 & 348 & FR/H   \\ 
        T0242   & 115 & \(\alpha/\beta\) &.27 &.30 & 0.89 & 358 & NF     \\
        \hline  \hline
        T0270-b & 125 & \(\alpha/\beta\) &.27 &.28 & 0.99 &  32 & --     \\ 
        T0270-a & 122 & \(\alpha/\beta\) &.25 &.27 & 0.80 &  47 & --     \\      
        T0272-b & 124 & \(\alpha/\beta\) & -- &.26 & 0.81 &  34 & FR/A   \\ 
        T0273   & 186 & \(\alpha/\beta\) &.22 &.24 & 0.98 & 189 & NF     \\ 
        \hline
      \end{tabular}\par}
  \end{table}

\begin{figure}{\par\centering
 \resizebox*{3.0in}{2.0in}{\includegraphics{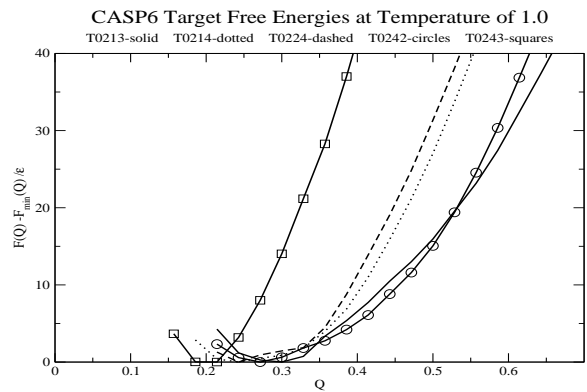}}
 \par}
  \caption{\label{fq_totalt0214t0243}Free Energy calculations for
         CASP6 targets T0213, T0214, T0224, T0242, and T0243.}
\end{figure}

  Calculating the free energy of several randomly chosen CASP6 targets
  in Fig.~\ref{fq_totalt0214t0243} provides us with probabilities of
  what we would have expected to see if more simulations has been
  performed during the CASP season.  We can estimate how
  many independent structures need to be seen at this temperature to
  sample the region 10 $k_BT$ greater than the minimum of the free
  energy.  We see roughly $e^{10}\approx 2*10^{4}$ independent sampled
  structures would be needed at a temperature of 1.0.  Target T0242 (PDB
  ID 2BLK) illustrates why the best structure we encountered had a Q
  score of 0.3. For this target, we sampled roughly 7000 different
  structures. To achieve a Q of 0.45, according to the free
  energy analysis we would need to increase our sampling by a factor 
  of 3.

  When extrapolating to lower temperatures, we see lower barriers to the
  folded state, and thus if sampling were more complete one would see
  better structures at these temperatures.  This further cooling would
  be a favorable strategy except that dynamic slowing due to the
  approach of the glass transition interferes, which occurs at a
  temperature of 0.9.  Naturally, it is best to sample just above the
  glass transition temperature, which can be approximately found from
  Q-Q correlation ($<Q(t)Q(t+\tau)>$) \cite{AllenTildesley}, and by
  using the Kolmogorov-Smirnov test to asses the independence of samples
  \cite{Eastwood2003}.  Table~\ref{casp6_likely} indicates what was the
  best structure we would be likely to see under such sampling
  conditions.  The differences between thermodynamically accessible
  structures and those that were sampled suggests that increased
  simulations would have improved the best structures sampled
  considerably.  The free energy of target T0243 (PDB ID not available) is
  significantly different due to its unusual architecture that contains
  a buried helix.

  \begin{table}
    \caption{\label{casp6_likely} Likely Quality of Structure Seen 
      at a Free  Energy of 10 CASP6 }
    {\centering \begin{tabular}{|c|c|c|c|c|} \hline 
        Target & PDB & length  & Probable Q & Sampled Q \\ \hline    
        T0213  & 1TE7 & 103&.43 &.32 \\ 
        T0214  & 1S04 & 110&.40 &.30 \\
        T0224  & 1RHX &  87&.39 &.38 \\ 
        T0242  & 2BLK & 123&.45 &.30 \\ 
        T0243  & ---  &  88&.28 &.32 \\ \hline
      \end{tabular}\par}
  \end{table}

\begin{figure}
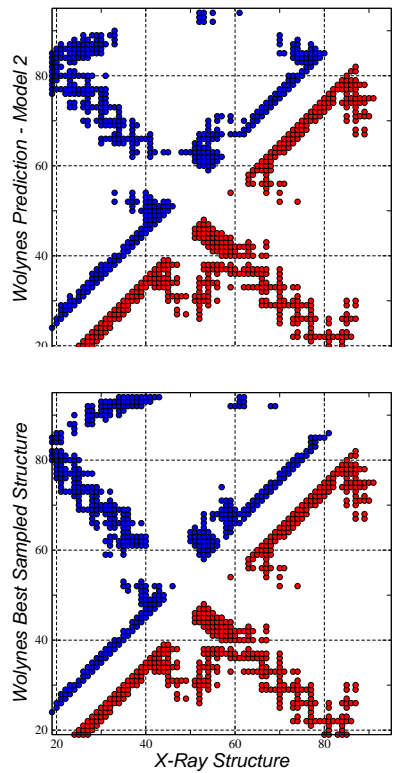
{\par\centering
 \resizebox*{2.0in}{2.0in}{\rotatebox{00}{\includegraphics{figures/contactbest_sub_t0201.eps}}}
 \hspace{.5in}
 \resizebox*{2.0in}{2.0in}{\rotatebox{00}{\includegraphics{figures/contactbest_samp_t0201.eps}}}
 \par}
 \caption{\label{contact_T0201}Contact maps for the best submitted
        (Q=.36) and the best sampled (Q=.44) structures for target T0201.}
 \end{figure}
\begin{figure}
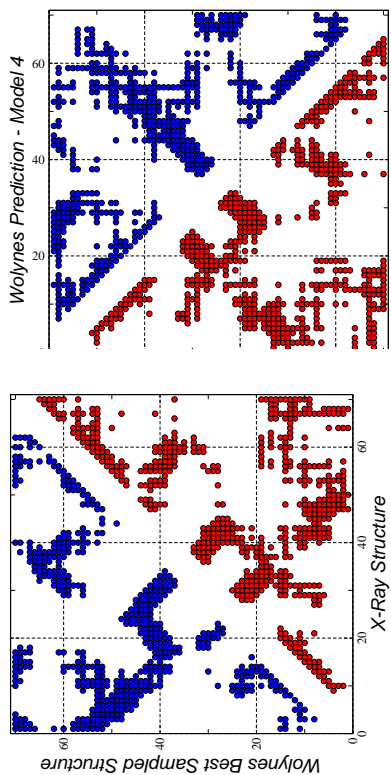
{\par\centering
 \resizebox*{2.0in}{2.0in}{\rotatebox{00}{\includegraphics{figures/contactbest_sub_t0281.eps}}}
 \hspace{.5in}
 \resizebox*{2.0in}{2.0in}{\rotatebox{90}{\includegraphics{figures/contactbest_samp_t0281.eps}}}
 \par}
 \caption{\label{contact_T0281}Contact maps for the best submitted
       (Q=.34) and the best sampled (Q=.48) structures for target T0281.}
 \end{figure}

  As in Fig.~\ref{casp5t0170contactmap}, we compare contact maps
  between the predictions and the experimentally resolved structure.
  Often contact maps give more insightful than superimposed structures
  especially when viewing in 2 dimensions.  We compare the submitted
  structures with the best structure encountered during our sampling to
  determine what aspect of folding are being captured by our energy
  functions. For a short target T0201 (PDB ID 1S12), we see that
  sometimes a small difference in the contact maps in
  Fig.~\ref{contact_T0201}, can greatly improve the quality of the
  prediction even though a large number of contacts are already correct.
  There was a larger fraction of incorrect contacts in our best
  submitted structure for target T0230 (PDB ID 1WCJ) than we would have
  seen in the best generated structure as shown in
  Fig.~\ref{contact_T0230}.  The incorrect parallel docking of the first
  two helices is largely resolved in the best sampled structure and the
  Q score improves considerably.  Similar analysis for target T0281 (PDB
  ID 1WHZ) shows incorrect long range contacts between the two otherwise
  properly oriented helices, and disordered intermediate interactions as
  in Fig.~\ref{contact_T0281}.  Again the best sampled structure has
  these problems largely resolved.

 \begin{figure}{\par\centering
 \resizebox*{2.0in}{2.0in}{\rotatebox{00}{\includegraphics{figures/contactbest_sub_t0230.eps}}}
 \hspace{.5in}
 \resizebox*{2.0in}{2.0in}{\rotatebox{00}{\includegraphics{figures/contactbest_samp_t0230.eps}}}
 \par}
     \caption{\label{contact_T0230}Contact maps for the best submitted
       (Q=.31) and the best sampled (Q=.42) structures for target T0230.}
 \end{figure}

\begin{figure}{\par\centering
       \resizebox*{3.5in}{3.0in}{\rotatebox{00}{\includegraphics{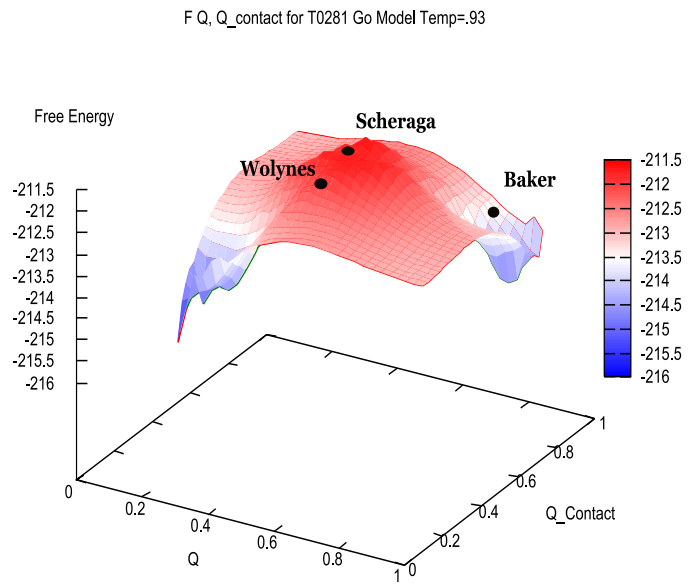}}}
       \par}
     \caption{\label{pfold_fig}G\=o Model Free Energy Surface with final
       prediction structures shown. The Pfold values for the three
       proteins are the Wolynes Group 0.07, Scheraga Group 0.02, and the
       Baker Group 0.97 with an error of +/- 0.1.}
   \end{figure}

  One amusing way to analyze predicted structures is to view the results
  of different structure prediction schemes as intermediates along a
  kinetic folding coordinate.  How far did the simulated annealing get in
  the folding pathway?  By mapping the likelihood of folding
  \cite{Shanknovich_1997} against its location on a folding free energy
  surface, we can assess how close the model structure is to the folded
  state in a kinetic sense.  The energy function for the kinetic
  modeling is a G\=o model \emph{i.e.} ideally non-frustrated energy
  function.  The difference between the G\=o model and the structure
  prediction energy functions is a measure of the quality of those
  structure prediction schemes. A pairwise additive G\=o model was
  created based on the native structure of the experimentally determined
  protein. As it has been discussed previously \cite{Eastwood00}, this
  G\=o model has both a polypeptide backbone energy terms that are the
  same as in the structure prediction energy function as described by
  Eq.~\ref{amh_back} and an interaction potential were the Gaussian
  interaction potential distances \(r_{ij}^{N}\) are determined by the
  native state formally described in Eq.~\ref{amh_go}.
  \begin{equation}
    \label{amh_go}
    E_{\text{G\=o}}=- \frac{\epsilon}{a} \sum_{i<j-3} \gamma_{\text{G\=o}}[x_{(|i-j|)}]\exp\left[-\frac{(r_{ij}-r_{ij}^{N})^2}{\sigma_{ij}^2}\right]
  \end{equation}
  The interactions are defined in this minimal model as residues with
  greater the three residues in sequence separation between \(
  C^{\alpha}-C^{\alpha}, C^{\alpha}-C^{\beta}, C^{\beta}-C^{\alpha},
  C^{\beta}-C^{\beta} \) atom pairs. The weights
  \(\gamma_{\text{G\=o}}\) or the depth of the Gaussian wells are set to
  (.177,.048,.430) in order to approximately divided the interaction
  energy equally between the different distance classes as defined in
  the original structure prediction energy function.  The width of the
  gaussians $\sigma_{ij}^2$ are defined by the sequence separation as
  before.  Notice that the G\=o Hamiltonian does not contain a summation
  over a set of memory structures as in the AMH, this is because all of
  the contacts in this definition of a G\=o model uses only the native
  state.  One hundred independent simulations of this G\=o energy
  function are performed starting with the best structure of three
  different structure prediction groups.  Pfold is then calculated by
  simply determining whether the simulation started from the model
  structure folds to the native structure or not.  The results in
  Fig.~\ref{pfold_fig} compare three minimalist models, one of which
  (the Baker Group) has undergone a further atomistic refinement. The
  minimalist models are only a few $k_BT$ from the barrier's peak, they
  only infrequently cross it.  It also suggests that a detailed less
  coarse grain sampling procedure maybe necessary for correctly
  assigning hydrophobic packing and hydrogen bonding patterns.

  \subsection*{The Next Generation in Structure Prediction}

\begin{figure}
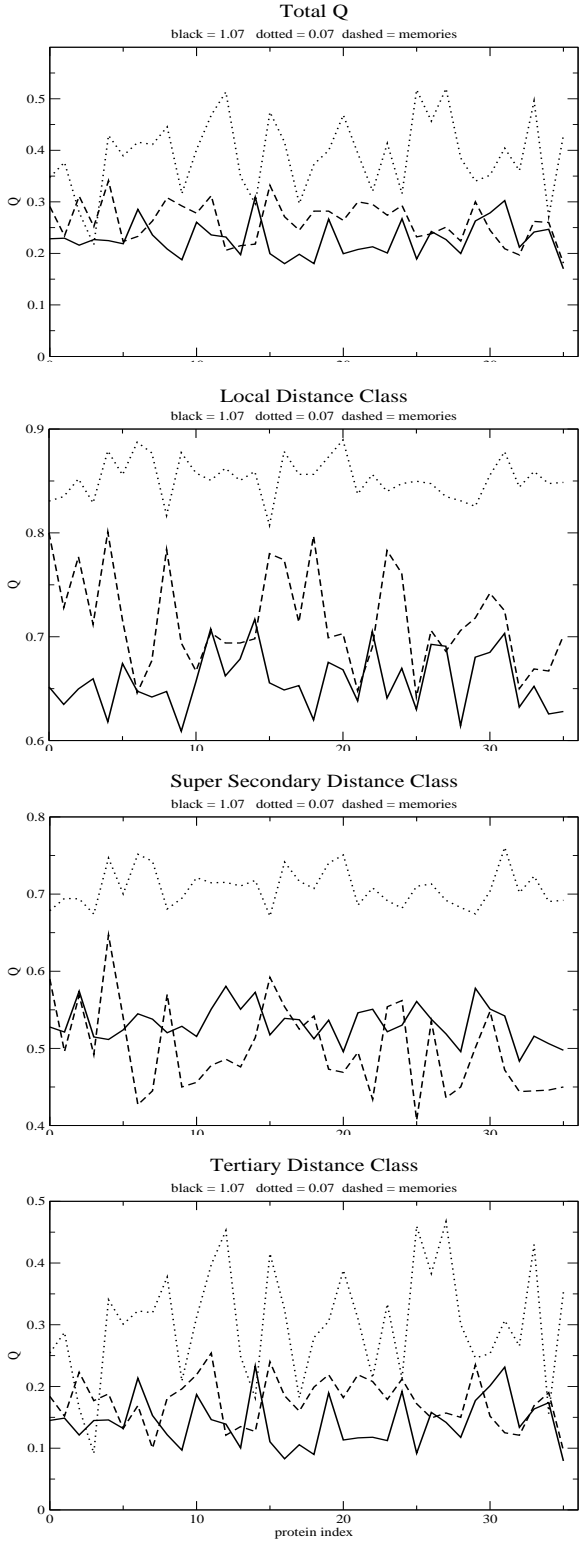
{\par\centering
 \resizebox*{3.0in}{2.0in}{\includegraphics{figures/256ba_qscore_bw.eps}}
 \resizebox*{3.0in}{2.0in}{\includegraphics{figures/256ba_qscore_short_bw.eps}}
 \resizebox*{3.0in}{2.0in}{\includegraphics{figures/256ba_qscore_medium_bw.eps}}
 \resizebox*{3.0in}{2.0in}{\includegraphics{figures/256ba_qscore_long_bw.eps}}
\par}
   \caption{\label{lh_256ba}These figures show the total Q, and the Q
     in the different distance classes between PDB structures,
     structures from a temperature of 1, and a temperature near zero
     for structures used as inputs to AMH simulations.  The
     lowest temperature show the largest improvement because they are
     fully collapsed.}
 \end{figure}

  Examining the contact maps of structures encountered during the CASP
  experiment, we observed that contacts between residues with a 
  large separation in
  sequence can be inaccurate, even when most of the contacts within a 12
  residues sequence separation are native like.  A different way of
  expressing this idea is the amount of funnelling is different within
  the different distance classes.  When comparing the quality of the
  intermediate range interactions in the sampled structures with the
  memories obtained with sequence analysis from the protein data bank, a
  dramatic increase of native-like interactions is seen as shown in
  Fig.~\ref{lh_256ba}.  While this was not used in the recent CASP
  exercise, we thought it would be interesting and straight forward to
  improve the prediction energy function by using these first generation
  results as better memory structures in the AMH.  Sequence to structure 
  alignments yield gap-less identity alignments thereby eliminating any 
  possibility of secondary structure registry shift irregularities.

  Different energy functions have been used to identify native like
  proteins from an ensemble of simulated structures.  Alternatively, one
  can rely on energy landscape ideas, and assume a mean field contact
  potential derived from the energy minima of the simulated energy
  function.  This approach has the additional advantage, that it does
  not rely on using a distinct energy function: one is simply seeing how
  close simulated annealing was to completely accessing the global
  minimum of the prediction energy function.  To select structures a
  pairwise Q denoted by a lower case q, is calculated between all of the
  ground state structures encountered in 200 independent simulations.

\begin{figure}
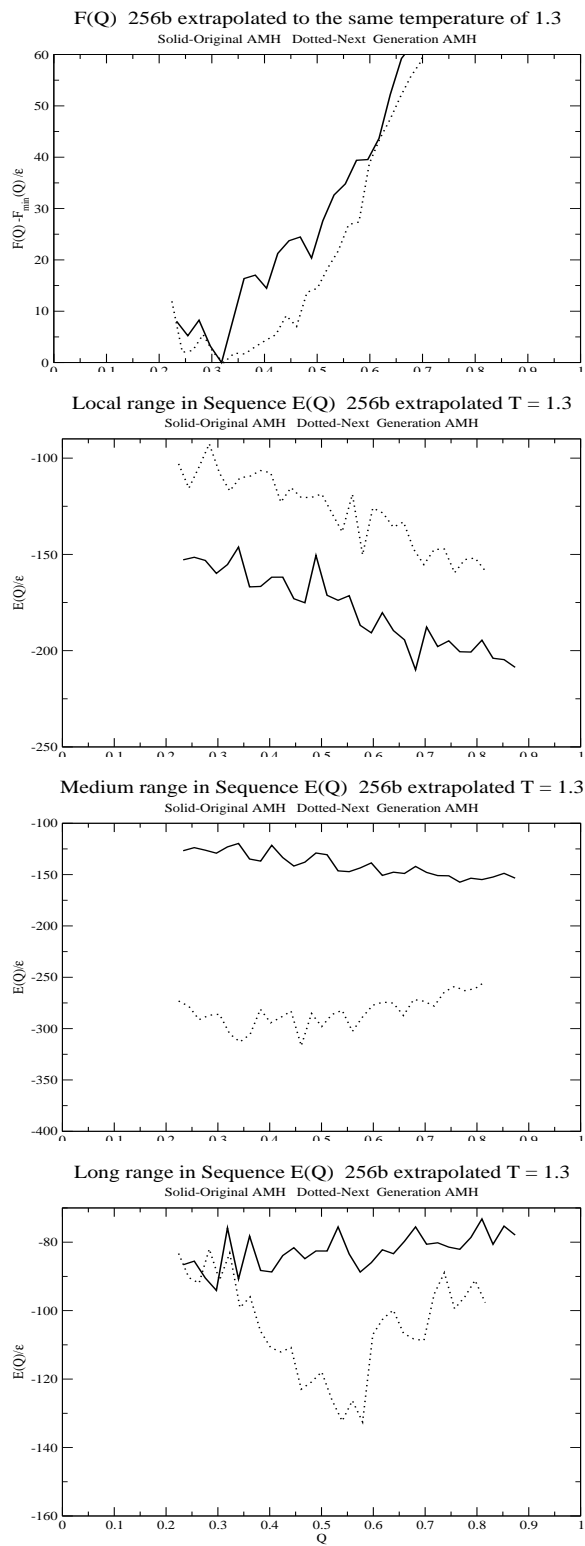
{\par\centering
\resizebox*{3.0in}{2.0in}{\includegraphics{figures/fq_bw_paper.eps}}
\resizebox*{3.0in}{2.0in}{\includegraphics{figures/eq_local_bw_paper.eps}}
\resizebox*{3.0in}{2.0in}{\includegraphics{figures/eq_ss_bw_paper.eps}}
\resizebox*{3.0in}{2.0in}{\includegraphics{figures/eq_tert_bw_paper.eps}}
\par}
   \caption{\label{amh_256ba}The free energy the two different energy
     functions for the protein 256B, shows roughly a 5-10 $k_BT$
     improvement for this protein.  The primary improvements are in the
     medium and long range distance classes.}
 \end{figure}

\begin{figure}
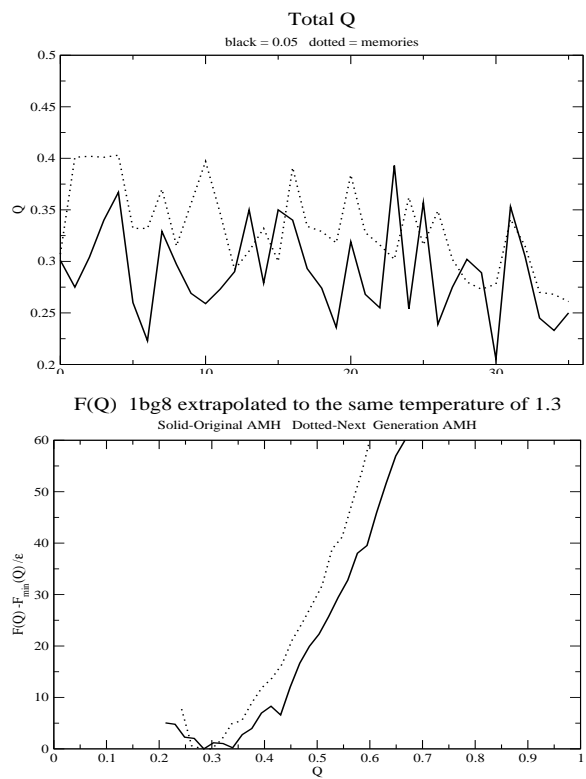

   \centering
\resizebox*{3.0in}{2.0in}{\includegraphics{figures/qscore_bw_1bg8.eps}}
\resizebox*{3.0in}{2.0in}{\includegraphics{figures/fq_bw_paper_1bg8.eps}}
\caption{\label{amh_1bg8a}The free energy the two different energy
     functions for the protein 1BG8 show little improvement.  The
     memories though show no enrichment in native contacts.}
 \end{figure}

  By dividing the inter-chain interactions under the same definitions as
  used in the energy function, the potential for improvements from such
  second generation structures over the original memories is
  considerable for protein 256B.  As seen in Fig.~\ref{lh_256ba}, the
  low temperature structure as identified by little q have an increased
  amount of native like contacts in all distance classes.  This style of
  analysis also suggests potential changes in the energy function.  The
  long distance in sequence interactions are also improved over that
  original memory used in the energy function.  In order to utilise this
  improvement the energy function in the distant interaction class was
  modified.  The original function used a multi-well contact potential,
  which does not use any information from the memory proteins.  For this
  third distance class the next generation energy function uses
  associative memory contacts much as was done before for modelling with
  homologues \cite{Koretke98}.  The energy function now takes the form
  \begin{equation}
    E_{\rm int}=-  \sum ^{c}_{3}\frac{\epsilon}{a_{c}}\sum ^{n}_{\mu }\sum ^{N}_{i<j} \gamma
    (P_{i}P_{j}P^{\mu }_{i'}P^{\mu }_{j'})\Theta
    (r_{ij}-r^{\mu }_{i'j'}). 
  \end{equation}
  The parameters for this new distance class are taken from the second
  distance class.  The total energy is defined over the set of memory
  structures as defined by Eq.~\ref{next_gen_units}
  \begin{equation}
    \label{next_gen_units}
    \epsilon=\frac{1}{36}\sum_{1}^{\mu}\frac{\left\vert E^{\rm model}_{\rm amh}\right\vert}{4N},
  \end{equation}
  instead of using the values taken from the optimisation.  Some next
  generation memory structures are more collapsed than the memory
  structures used in initial round of simulation.  Furthermore the
  scaling is changed from the initial round of simulation's 1:1:1
  scaling amongst the three different (local, super-secondary, tertiary)
  distance classes to 1.5:0.5:1 in an effort to approximate the equal
  division of energy in each distance class. To examine the equilibrium 
  properties of this energy function, we need
  to estimate the glass transition temperature.  As previously explored
  \cite{Eastwood2003}, we use the Kolmogorov-Smirnov test to determine
  if two independent simulations have been sampled from the same
  equilibrium distribution.  This test ensures that
  simulations are equilibrated.
 \begin{figure}
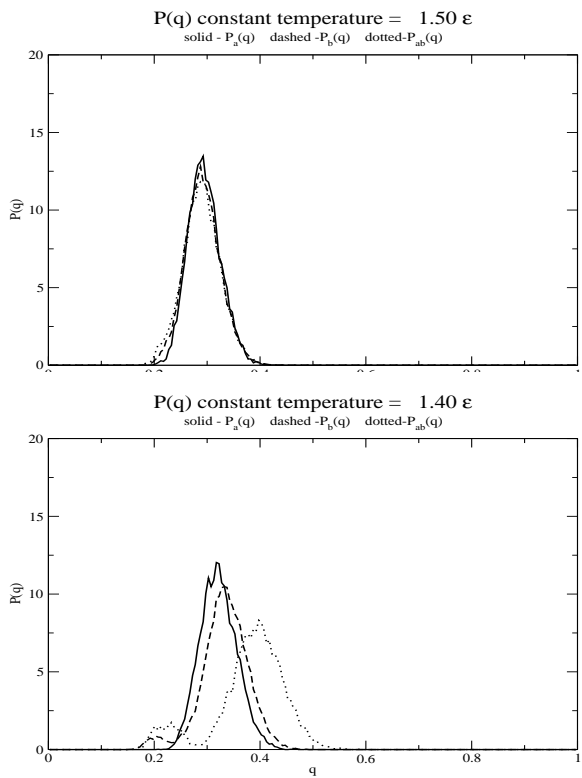

   \centering
   \resizebox*{3.0in}{2.0in}{\includegraphics{figures/Pq_black_white_15.eps}}
   \hspace{.5in}
   \resizebox*{3.0in}{2.0in}{\includegraphics{figures/Pq_black_white_14.eps}}
   \caption{\label{ks_256ba}Kolmogorov-Smirnov test shows the constant
     temperature simulation falling out of equilibrium at a lower
     temperature of 1.4. The different probability distributions of
     structures between two independent simulations is no longer the
     same.}
 \end{figure}
 Once the glass transition temperature ($T_{K}$) is estimated using the
 Kolmogorov-Smirnov test, we can use standard techniques to
 quantify the equilibrium properties of different energy functions.
 The proteins we used for study of the next generation AMH strategy are
 cytochrome B562 (PDB ID 256b), HDEA (PDB ID 1BG8), because they are
 both of moderate size and one of them (1BG8) was not in the training
 set of proteins that optimized the original energy function. An
 additional advantage of this choice is these proteins have different
 fold types. According to CATH \cite{FPearl} HDEA belongs to the
 orthogonal bundle architecture, while cytochrome B562 represents an
 up-down bundle.  Using umbrella sampling combined with the weighted
 histogramming method, we are able to sample parts of phase space that
 would rarely be encountered during a simulation \cite{kong96}.  When
 using memories with a larger number of native contacts, we see
 improved free energy and energy profiles as shown in
 Fig.~\ref{amh_256ba}. 
 This is even more impressive when we consider
 this energy function has not yet been properly optimised for this new
 hamiltonian.  For the other target, the results are also not
 surprising.  In this case the next generation memories used to simulate
 this protein were not of greater structural quality than the
 initial set.  Thus a very similar free energy profile was generated as
 seen in Fig.~\ref{amh_1bg8a}.  Our use of q as an order
 parameter was successful in identifying the high Q protein for the
 256B example.  This is due to the highly funnelled characteristic of
 the first generation energy function.  The original energy function
 for 1BG8 is not as funnelled so therefore there is poorer enrichment by
 scanning with little q. This limitation could be over come by increasing 
 the amount of sampling of structures in the first generation simulations. 
 More simulations would guarantee better structure
 as was demonstrated during the CASP5 exercise. This difference in the 
 enrichment could be anticipated by using the Kolmogorov-Smirnov measure
 to differentiate the distribution of the q values encountered between
 structures derived from simulation and the protein databank.

 \section*{Conclusion}

 These case studies from our participation in the CASP experiment only
 provide a snap shot of our group's prediction schemes.  It produces a
 series of lessons for us and we hope for others.  In the future, a
 more balanced efforts between the sampling and selection of structures
 from that ensemble would appear to be desirable.  More efforts in
 selection would have clearly improved the results submitted in CASP6.
 While it is was computationally impractical to quench all of the
 structures simulated during the prediction season, the comparison of
 the contact maps demonstrated further that tempering of the structure
 would have improved intermediate range ordering.  Using preliminary
 structures as input to a next generation of AMH modelling improves the
 quality of the prediction results.  While these results may initially
 appear to be model or energy function specific, we feel that any
 algorithm that uses structures as an input would benefit from similar
 next generation approaches.

 \section*{Acknowledgments}
 The authors thank Joe Hegler, Zaida Luthey-Schulten, Garegin Papoian,
 and Marcio Von Muhlen for their key roles in developing codes used in
 this study and for many helpful discussions over the years.  The
 efforts of P.G.W. are supported through the National Institutes of
 Health Grant 5RO1GM44557.  Computing resources were supplied by the
 Center for Theoretical Biological Physics through National Science
 Foundation Grants PHY0216576 and PHY0225630.

\bibliographystyle{ieeetr}

\begin{thebibliography}{10}

\bibitem{Moult}
Moult,~J.;\ \ Fidelis,~K.;\ \ Zemla,~A.;\ \ Hubbard,~T. \textit{Proteins}
  \textbf{2003,} \textsl{53 Suppl 6,} 334-339.

\bibitem{GoldsteinRA-AMH-92}
Goldstein,~R.~A.;\ \ Luthey-Schulten,~Z.~A.;\ \ Wolynes,~P.~G. \textit{Proc
  Natl Acad Sci USA} \textbf{1992,} \textsl{89,} 4918-4922.

\bibitem{BryngelsonJD87}
Bryngelson,~J.~D.;\ \ Wolynes,~P.~G. \textit{Proc Natl Acad Sci USA}
  \textbf{1987,} \textsl{84,} 7524-7528.

\bibitem{AnfinsenCB73}
Anfinsen,~C.~B. \textit{Science} \textbf{1973,} \textsl{181,} 223-230.

\bibitem{GoN83}
G{\={o}},~N. \textit{Annu Rev Biophys and Bioeng} \textbf{1983,} \textsl{12,}
  183-210.

\bibitem{Koga_Takada01}
Koga,~N.;\ \ Takada,~S. \textit{J Mol Biol} \textbf{2001,} \textsl{313,}
  171-180.

\bibitem{portman98}
Portman,~J.~J.;\ \ Takada,~S.;\ \ Wolynes,~P.~G. \textit{Phys Rev Lett}
  \textbf{1998,} \textsl{81,} 5237--5240.

\bibitem{Wales}
Wales,~D. \textit{Energy Landscapes;} Cambridge University Press: Cambridge,
  UK, 2003.

\bibitem{Wheelan_etal00}
Wheelan,~S.~J.;\ \ Marchler-Bauer,~A.;\ \ Bryant,~S.~H. \textit{Bioinformatics}
  \textbf{2000,} \textsl{16,} 613-618.

\bibitem{Heringa02}
George,~R.~A.;\ \ Heringa,~J. \textit{J Mol Biol} \textbf{2002,} \textsl{316,}
  839-851.

\bibitem{Rigden02}
Rigden,~D.~J. \textit{Protein Eng} \textbf{2002,} \textsl{15,} 65-77.

\bibitem{Hardin2002ab}
Hardin,~C.;\ \ Eastwood,~M.;\ \ Prentiss,~M.;\ \ Luthey-Schulten,~Z.;\ \
  Wolynes,~P.~G. \textit{Proc. Nat. Acad. Sci. U.S.A.} \textbf{2002,}
  \textsl{100,} 1679-1684.

\bibitem{Eastwood00}
Eastwood,~M.~P.;\ \ Hardin,~C.;\ \ Luthey-Schulten,~Z.;\ \ Wolynes,~P.~G.
  \textit{IBM Systems Research} \textbf{2001,} \textsl{45,} 475-497.

\bibitem{KoretkeKK96}
Koretke,~K.~K.;\ \ Luthey-Schulten,~Z.;\ \ Wolynes,~P.~G. \textit{Protein Sci}
  \textbf{1996,} \textsl{5,} 1043-1059.

\bibitem{Berman}
Berman,~H.~M.;\ \ Westbrook,~J.;\ \ Feng,~Z.;\ \ Gilliland,~G.;\ \
  Bhat,~T.~N.;\ \ Weissig,~H.;\ \ Shindyalov,~I.~N.;\ \ Bourne,~P.~E.
  \textit{Nucl. Acids Res.} \textbf{2000,} \textsl{28,} 235-242.

\bibitem{Bourne98}
Shindyalov,~I.;\ \ Bourne,~P. \textit{Protein Engineering} \textbf{1998,}
  \textsl{11,} 739-747.

\bibitem{FriedrichsMS89}
Friedrichs,~M.~S.;\ \ Wolynes,~P.~G. \textit{Science} \textbf{1989,}
  \textsl{246,} 371-373.

\bibitem{FriedrichsMS90}
Friedrichs,~M.;\ \ Wolynes,~P.~G. \textit{Tet Comp Meth} \textbf{1990,}
  \textsl{3,} 175.

\bibitem{FriedrichsMS91}
Friedrichs,~M.~S.;\ \ Goldstein,~R.~A.;\ \ Wolynes,~P.~G. \textit{J Mol Biol}
  \textbf{1991,} \textsl{222,} 1013-1034.

\bibitem{Hardin00}
Hardin,~C.;\ \ Eastwood,~M.;\ \ Luthey-Schulten,~Z.;\ \ Wolynes,~P.~G.
  \textit{Proc Natl Acad Sci USA} \textbf{2000,} \textsl{97,} 14235-14240.

\bibitem{GoldsteinRA92}
Goldstein,~R.;\ \ Luthey-Schulten,~Z.~A.;\ \ Wolynes,~P.~G. \textit{Proc Natl
  Acad Sci USA} \textbf{1992,} \textsl{89,} 9029-9033.

\bibitem{Hopfield_1982}
Hopfield,~J.~J. \textit{Proc Natl Acad Sci USA} \textbf{1982,} \textsl{79,}
  2554-2558.

\bibitem{Ryckaert77}
Ryckaert,~J.;\ \ Ciccotti,~G.;\ \ Berendsen,~H. \textit{J Comput Phys}
  \textbf{1977,} \textsl{23,} 327-341.

\bibitem{Rama}
Ramachandran,~G.;\ \ Sasisekharan,~V. \textit{Adv Protein Chem} \textbf{1968,}
  \textsl{23,} 283-438.

\bibitem{Papoian04pnas}
Papoian,~G.~A.;\ \ Ulander,~J.;\ \ Eastwood,~M.~P.;\ \ Luthey-Schulten,~Z.;\ \
  Wolynes,~P.~G. \textit{Proc Natl Acad Sci U S A} \textbf{2004,} \textsl{101,}
  3352-3357.

\bibitem{Papoian03biopoly}
Papoian,~G.~A.;\ \ Wolynes,~P.~G. \textit{Biopolymers} \textbf{2003,}
  \textsl{68,} 333-349.

\bibitem{Papoian03jacs}
Papoian,~G.~A.;\ \ Ulander,~J.;\ \ Wolynes,~P.~G. \textit{J Am Chem Soc}
  \textbf{2003,} \textsl{125,} 9170-9178.

\bibitem{maxfield79}
Maxfield,~F.~R.;\ \ Scheraga,~H.~A. \textit{Biochemistry} \textbf{1979,}
  \textsl{18,} 697--704.

\bibitem{finkelstein98}
Finkelstein,~A.~V. \textit{Phys Rev Lett} \textbf{1998,} \textsl{80,}
  4823-4825.

\bibitem{Altschul1997}
Altschul,~S.;\ \ Madden,~T.;\ \ Schaffer,~A.;\ \ Zhang,~J.;\ \ Zhang,~Z.;\ \
  Miller,~W.;\ \ Lipman,~D. \textit{Nucl. Acids Res.} \textbf{1997,}
  \textsl{25,} 3389-3402.

\bibitem{bioperl}
Stajich,~J.~E. \textit{et al.}\  \textit{Genome Res.} \textbf{2002,}
  \textsl{12,} 1611-1618.

\bibitem{CLUSTAL}
Thompson,~J.;\ \ Higgins,~D.;\ \ Gibson,~T. \textit{Nucl. Acids Res.}
  \textbf{1994,} \textsl{22,} 4673-4680.

\bibitem{Zhang_etal03}
Zhang,~Y.;\ \ Kolinski,~A.;\ \ Skolnick,~J. \textit{Biophys J} \textbf{2003,}
  \textsl{85,} 1145-1164.

\bibitem{Hardin02a}
Hardin,~C.;\ \ Eastwood,~M.;\ \ Prentiss,~M.;\ \ Luthey-Schulten,~Z.;\ \
  Wolynes,~P.~G. \textit{J Comput Chem} \textbf{2002,} \textsl{23,} 138-146.

\bibitem{BONN2001b}
Bonneau,~R.;\ \ Tsai,~J.;\ \ Ruczinski,~I.;\ \ Chivian,~D.;\ \ Rohl,~C.;\ \
  Strauss,~C. E.~M.;\ \ Baker,~D. \textit{Proteins} \textbf{2001,}
  \textsl{Suppl 5,} 119-126.

\bibitem{zhou:zhou04}
Zhou,~H.;\ \ Zhou,~Y. \textit{Proteins} \textbf{2004,} \textsl{54,} 315-322.

\bibitem{kussell:168101}
Kussell,~E.;\ \ Shakhnovich,~E.~I. \textit{Phys Rev Lett} \textbf{2002,}
  \textsl{89,} 168101.

\bibitem{Baker97}
Simons,~K.;\ \ Kooperberg,~C.;\ \ Huang,~E.;\ \ Baker,~D. \textit{J. Mol.
  Biol.} \textbf{1997,} \textsl{268,} 209-225.

\bibitem{Betancourt:2004}
Betancourt,~M.;\ \ Skolnick,~J. \textit{J Mol Biol} \textbf{2004,} \textsl{2,}
  635-649.

\bibitem{SavenJG96}
Saven,~J.~G.;\ \ Wolynes,~P.~G. \textit{J. Mol. Biol.} \textbf{1996,}
  \textsl{257,} 199-216.

\bibitem{AllenTildesley}
Allen,~M.~P.;\ \ Tildesley,~D.~J. \textit{Computer Simulation of Liquids;}
  Clarendon Press: New York, NY, USA, 1987.

\bibitem{Eastwood2003}
Eastwood,~M.;\ \ Hardin,~C.;\ \ Luthey-Schulten,~Z.;\ \ Wolynes,~P.~G.
  \textit{J Chem Phys} \textbf{2003,} \textsl{118,} 8500-8512.

\bibitem{Shanknovich_1997}
Du,~R.;\ \ Pande,~V.;\ \ A.Y.,~G.;\ \ Shakhnovich,~E.~I. \textit{J Chem Phys}
  \textbf{1997,} \textsl{108,} 334-350.

\bibitem{Koretke98}
Koretke,~K.~K.;\ \ Luthey-Schulten,~Z.;\ \ Wolynes,~P.~G. \textit{Proc Natl
  Acad Sci USA} \textbf{1998,} \textsl{95,} 2932-2937.

\bibitem{FPearl}
Pearl,~F. \textit{et al.}\  \textit{Nucl. Acids Res.} \textbf{2005,}
  \textsl{33,} D247-251.

\bibitem{kong96}
Kong,~X.;\ \ {Brooks III},~C.~L. \textit{J Chem Phys} \textbf{1996,}
  \textsl{105,} 2414--2423.

\end{thebibliography}

\providecommand{\refin}[1]{\\ \textbf{Referenced in:} #1}

\end{document}